%% file: Main.tex
\documentclass[preprint,12pt]{elsarticle}

\usepackage{graphicx}
\graphicspath{{Figures/}}
\usepackage{amssymb}

\usepackage{lineno}

\usepackage{mdframed}
\usepackage{listings}
\usepackage{url}
\usepackage{amsmath,amsfonts,amssymb}
\usepackage[ruled,vlined]{algorithm2e}

\definecolor{codegreen}{rgb}{0,0.6,0}
\definecolor{codegray}{rgb}{0.5,0.5,0.5}
\definecolor{codepurple}{rgb}{0.58,0,0.82}
\definecolor{backcolour}{rgb}{0.95,0.95,0.92}

\lstdefinestyle{mystyle}{
    backgroundcolor=\color{backcolour},   
    commentstyle=\color{codegreen},
    keywordstyle=\color{magenta},
    numberstyle=\tiny\color{codegray},
    stringstyle=\color{codepurple},
    basicstyle=\ttfamily\footnotesize,
    breakatwhitespace=false,         
    breaklines=true,                 
    captionpos=b,                    
    keepspaces=true,                 
    numbers=right,                    
    numbersep=5pt,                  
    showspaces=false,                
    showstringspaces=false,
    showtabs=false,                  
    tabsize=2
}

\journal{Elsevier}
\usepackage{subfigure}

\begin{document}

\begin{frontmatter}


\title{PythonFOAM: In-situ data analyses with OpenFOAM and Python}



\author[label1]{Romit Maulik}
\ead{rmaulik@anl.gov}
\author[label2,label3]{Dimitrios K. Fytanidis}
\author[label1]{Bethany Lusch}
\author[label1]{Venkatram Vishwanath}
\author[label2]{Saumil Patel}

\address[label1]{Argonne Leadership Computing Facility, Bldg. 240, Argonne National Laboratory, Lemont, IL-60439}

\address[label2]{Computational Science Division, Bldg. 240, Argonne National Laboratory, Lemont, IL-60439}

\address[label3]{Department of Civil and Environmental Engineering, Ven Te Chow Hydrosystems Laboratory, University of Illinois at Urbana-Champaign, IL-61801}

\begin{abstract}
We outline the development of a general-purpose Python-based data analysis tool for OpenFOAM. Our implementation relies on the construction of OpenFOAM applications that have bindings to data analysis libraries in Python. Double precision data in OpenFOAM is cast to a NumPy array using the NumPy C-API and Python modules may then be used for arbitrary data analysis and manipulation on flow-field information. We highlight how the proposed wrapper may be used for an in-situ online singular value decomposition (SVD) implemented in Python and accessed from the OpenFOAM solver PimpleFOAM. Here, `in-situ' refers to a programming paradigm that allows for a concurrent computation of the data analysis on the same computational resources utilized for the partial differential equation solver. In addition, to demonstrate data-parallel analyses, we deploy a distributed SVD, which collects snapshot data across the ranks of a distributed simulation to compute the global left singular vectors. Crucially, both OpenFOAM and Python share the same message passing interface (MPI) communicator for this deployment which allows Python objects and functions to exchange NumPy arrays across ranks. Subsequently, we provide scaling assessments of this distributed SVD on multiple nodes of Intel Broadwell and KNL architectures for canonical test cases such as the large eddy simulations of a backward facing step and a channel flow at friction Reynolds number of 395. Finally, we demonstrate the deployment of a deep neural network for compressing the flow-field information using an autoencoder to demonstrate an ability to use state-of-the-art machine learning tools in the Python ecosystem.
\end{abstract}


\end{frontmatter}




\section{Introduction}
\label{S:1}


\input{Source/Introduction.tex}

\section{PythonFOAM: Calling Python modules in OpenFOAM}

\subsection{Python embedding}

\input{Source/Python_Embedding.tex}

\subsection{Compiling and linking}

\input{Source/Compiling_Linking.tex}

\section{Algorithms}

\input{Source/Algorithms.tex}

\section{Experiments}

\input{Source/Experiments}

\section{Conclusions and future work}

\input{Source/Conclusions}

\section{CRediT author statement}

Romit Maulik: Conceptualization, methodology, software, formal analysis, writing - original draft, writing - reviewing \& editing. Dimitrios K. Fytanidis: analysis, visualization, reviewing \&  editing, Bethany Lusch: methodology, supervision, analysis, Venkatram Vishwanath: methodology, supervision, analysis, Saumil Patel: analysis.

\section*{Acknowledgements}

The authors thank Zhu Wang and Traian Iliescu for helpful discussions related to the distributed singular value decomposition. This material is based upon work supported by the U.S. Department of Energy (DOE), Office of Science, Office of Advanced Scientific Computing Research, under Contract DE-AC02-06CH11357. This research was funded in part and used resources of the Argonne Leadership Computing Facility, which is a DOE Office of Science User Facility supported under Contract DE-AC02-06CH11357. RM acknowledges support from the Margaret Butler Fellowship at the Argonne Leadership Computing Facility. This paper describes objective technical results and analysis. Any subjective views or opinions that might be expressed in the paper do not necessarily represent the views of the U.S. DOE or the United States Government.

\begin{mdframed}
    The submitted manuscript has been created by UChicago Argonne, LLC, Operator of Argonne National Laboratory (``Argonne”). Argonne, a U.S. Department of Energy Office of Science laboratory, is operated under Contract No. DE-AC02-06CH11357. The U.S. Government retains for itself, and others acting on its behalf, a paid-up nonexclusive, irrevocable worldwide license in said article to reproduce, prepare derivative works, distribute copies to the public, and perform publicly and display publicly, by or on behalf of the Government. The Department of Energy will provide public access to these results of federally sponsored research in accordance with the DOE Public Access Plan (http://energy.gov/downloads/doe-public-access-plan).
\end{mdframed}

\bibliographystyle{elsarticle-num-names}
\bibliography{Source/sample.bib}

\appendix
\section*{Appendix A}
\label{Appendix_A}

\input{Source/Appendix}

\end{document}

%% file: Source/Introduction.tex
For most large-scale computational studies, users are frequently required to make difficult decisions with respect to how often simulation data must be exported to storage. This mainly pertains to the limitations of input-output (I/O) bandwidth and the desire to limit the ratio of compute time to I/O time. \emph{In-situ} data analyses and machine learning strategies promise an alternate route to reducing this ratio and consequently reduce the time to solution for the simulation workflow. This allows for the user to export the post-analysis quantities of interest to storage (as compressed data, visualization, or discovered models). This also allows a user to have greater control over the temporal resolution of the analyses; for instance passing checkpoint data to a machine learning algorithm would not be limited by a storage bottleneck. In essence, in-situ algorithms and software provide a promising avenue to bypass the limitations of the classical pre-processing, simulation, and post-processing workflow, particularly as modern solvers start leveraging exascale infrastructure \cite{klasky2011situ}.

In this article we focus on OpenFOAM \cite{weller1998tensorial}, a well-established open-source finite-volume code for computational fluid dynamics. OpenFOAM has been used across industry and academia for a diverse range of applications such as coastal engineering \cite{higuera2013simulating}, wave-structure interaction \cite{chen2014numerical}, multiphase heat transfer \cite{kunkelmann2009cfd}, turbulent separated flows \cite{lysenko2013modeling}, cavitation \cite{bensow2010simulating}, non-Newtonian flows \cite{favero2010viscoelastic}, design optimization \cite{he2018aerodynamic}, rarefied gas dynamics \cite{white2018dsmcfoam+}, large eddy simulations \cite{tabor2010inlet,laurila2019analysis} etc. Therefore, any tool for in-situ data analysis in OpenFOAM has the potential to be highly impactful across many domains. Currently, there are no packages that can provide a robust, easy-to-use, and flexible interface between Python and OpenFOAM. Instead, most studies have relied on application-specific combinations of OpenFOAM and data analysis tools that do not readily generalize to arbitrary problems.

Geneva and Zabaras \cite{geneva2019quantifying} embed a neural network model into OpenFOAM 4.1 using the PyTorch C backend. However, this procedure requires the installation of additional software packages such as ONNX and Caffe2 that may cause issues with dependencies. In addition, Caffe2 is deprecated (to be subsumed into PyTorch), and future incorporation of PyTorch models into OpenFOAM through this route is unclear. Another tool under active development is the Fortran-Keras Bridge (FKB) \cite{ott2020fortran}, which successfully couples densely connected neural networks to Fortran simulation codes. However, FKB is yet to support more complicated architectures such as convolutional neural networks, and development revolves around the programming of neural network subroutines in Fortran before a Keras model can be imported. A similar tool, MagmaDNN \cite{nichols2019magmadnn}, has been developed in C++, with an emphasis on neural architectures. While the neural networks supported in this package are much more extensive, off-nominal architectures (for instance generative models for density estimation) would require specialized development. Another recent development for in-situ visualization and analysis has been demonstrated for Nek5000 \cite{bernardoni2018situ} where the SENSEI framework has been used to couple simulations with VisIt \cite{childs2012visit}. While this capability allows for visualization at extreme scales due to the excellent scaling of the higher-order spectral methods implemented in Nek5000, the primary focus of VisIt is to visualize data and to perform rudimentary analyses. \textcolor{black}{Another popular software for extreme scale in-situ visualizations and data-analyses is provided by Paraview Catalyst \cite{ayachit2015paraview}. However, Catalyst is predominantly set up as a data visualization tool and does not possess advanced state-of-the-art analysis techniques that may require deep learning technologies. We note that the Python/C coupling proposed in this article has the ability to complement Catalyst since the latter can be interfaced with NumPy arrays.} To build on the successes of existing technologies, the current authors have previously demonstrated how the C API of TensorFlow may be utilized from within OpenFOAM for in-situ surrogate modeling applications \cite{maulik2020turbulent,maulik2021deploying}. While these tools have removed the limitations of neural network development in C++ or Fortran, and also allows for the utilization of a wider choice of data analysis functions, the user is limited to data-driven analyses in TensorFlow alone. Another example, PAR-RL, has been developed along the same lines to allow for deep reinforcement learning integration with OpenFOAM \cite{pawar2021distributed} but relies on a filesystem-based information exchange between the reinforcement learning agent and the numerical simulation being controlled. Therefore, in this project, we demonstrate the usage of the Python/C++ interoperability for tighter integration of data-science and computational fluid dynamics capabilities.

Our goal, through this research, is to address the inflexible nature of data analysis tools for computational fluid dynamics codes by using the Python/C++ API for integrating data-science capability with to OpenFOAM. These bindings may then be used for a broad range of data analyses in concurrence with the simulation. For example, we shall demonstrate how one may leverage NumPy \cite{harris2020array} linear algebra capabilities for rapid deployment of data analysis routines. This is achieved by enabling a data interface from OpenFOAM data structures to functions that can handle NumPy arrays. The end result is a pipeline that can allow for the application of arbitrary functions on the simulation data, now represented as NumPy arrays. We demonstrate our wrapper through the implementation of an online singular value decomposition (SVD) \cite{levy1998sequential} that is used to extract coherent structures from the flow field without any storage I/O. We also demonstrate the calculation of a distributed SVD using the approximate partitioned method of snapshots \cite{wang2016approximate} where data from multiple ranks is accessed by the MPI4PY library in Python \cite{dalcin2005mpi}. We outline results from our deployment on multiple ranks of Intel Broadwell and Knights-Landing CPUs and assess strong scaling for moderate sized problems. \textcolor{black}{Finally, we also demonstrate how our data science integration technique enables the use of state-of-the-art tools in machine learning such as TensorFlow \cite{tensorflow2015} by including an example of nonlinear compression using an autoencoder (a deep neural network with a bottleneck architecture). Note that interoperability with other Python libraries such as scikit-learn and PyTorch are also possible \cite{NEURIPS2019_9015}.  The experiments presented in this paper assume a blocking usage of the data analyses routines in Python, wherein the numerical solver is paused while the same resource is utilized for the Python function execution. While this may be a limitation in terms of the optimal usage of computational resources in certain scenarios, and will be addressed in future augmentations to this work, we address the issue of minimizing read and write operations to and from the disk. }

%% file: Source/Python_Embedding.tex
In this section, we introduce how one can call Python from OpenFOAM. We utilize the Python/C API \footnote{\url{https://docs.Python.org/3/c-api/intro.html}} which conveniently allows for C++ code to import Python modules with generic class objects and functions within them. While the API is more commonly used for \emph{extending} Python capabilities with functions written in C or C++, it may also be used for calling Python within a larger application, which is generally referred to as \emph{embedding} Python in an application. \textcolor{black}{We note that there are alternative packages available for establishing the bridge between Python and C++ such as Pybind and Boost and the overall coupling strategy we have demonstrated here may be replicated with them as well.}

In the following, we outline how one may embed Python in OpenFOAM. Specifically, we highlight how a pre-existing solver (such as pimpleFOAM) may be readily modified to exchange information with a Python interpreter. The first step in this process is for OpenFOAM to initialize a Python interpreter that must remain live for the entire duration of the simulation. This is accomplished by using \texttt{Py\_Initialize()}. Following this, the Python interpreter may be interacted with from within OpenFOAM in a manner similar to the command line using \texttt{PyRun\_SimpleString()}, where the argument to this function is the Python code one wishes to execute. For instance, to ensure that Python is able to discover modules in the current working directory, it is common to execute lines 3 and 4 in Listing \ref{LST1}. In practice, this functionality by itself is insufficient for arbitrary interaction of data, visualization, and compute between C++ and Python. Therefore, one needs to utilize Python modules and their functions to interact with OpenFOAM data structures directly. Information about modules and function names are stored in pointers to \texttt{PyObject}. Similarly, data that is sent to (or received from) Python is stored in these pointers as well. 

Lines 13 and 14 in Listing \ref{LST1}, for example, show a pointer \texttt{pName} that stores the name of the module we wish to import (\texttt{python\_module}) and a pointer \texttt{pModule} that stores the imported module respectively. Modules are imported using \texttt{PyUnicode\_DecodeFSDefault()} with the name of Python module as the argument (which should be present in the current working directory). Here a module from a Python file, \texttt{python\_module.py}, present in the OpenFOAM case directory will be loaded once at the start of the solver. Similarly, line 17 details how \texttt{PyObject\_GetAttrString()} may be used to import a function (in this case \texttt{python\_func} from \texttt{python\_module}).  Finally, any arguments that may be needed for calling must be stored in a tuple as given in Line 20, before they may be passed through the API to the Python interpreter. Before proceeding, we note that an additional command, \texttt{import\_array1(-1)}, is utilized to initialize the ability to use NumPy data structures from within OpenFOAM. NumPy allows for flexible and efficient data analysis within Python and can interface with a vast number of specialized tools in the Python ecosystem. \texttt{Py\_DECREF()} is used for memory deallocation after \texttt{pName} and \texttt{pModule} use is completed.

\begin{lstlisting}[language=C++,style=mystyle,caption=Enabling interoperability between C++ and Python. This listing details how one may load arbitrary Python modules and functions from C++ for downstream use in computation and is modified from the Python C/API manual.,label=LST1]
/* Initialize Python and add current directory to path*/
Py_Initialize();
PyRun_SimpleString("import sys");
PyRun_SimpleString("sys.path.append(\".\")");

/*Python API datastructures*/
PyObject *array_2d, *rank_val;

/* initialize numpy array library */
import_array1(-1); 

/*Import the "python_module.py" module from the current directory*/
PyObject *pName = PyUnicode_DecodeFSDefault("python_module"); // Python filename
PyObject *pModule = PyImport_Import(pName);

/*Import the function "python_func"*/
PyObject *python_func = PyObject_GetAttrString(pModule, "python_func");

/*The tuple contains the items that need to be passed to the "python_func" function*/
PyObject *python_func_args = PyTuple_New(2);

/*Numpy datastructure to obtain return value from python method*/
PyArrayObject *pValue;

/*Release memory*/
Py_DECREF(pName);
Py_DECREF(pModule);
\end{lstlisting}

Following the initialization of the Python interpreter and the loading of modules and functions, Listing \ref{LST2} outlines how one may pass data from OpenFOAM to Python via the Python/C and NumPy/C APIs. Here we show how a generic Python function may be called from OpenFOAM while passing data from OpenFOAM to it, and how its return value may be stored in an OpenFOAM compatible data structure. We first retrieve data from OpenFOAM's \texttt{volVectorField} data structure and store it in an intermediate double precision array as shown between lines 1 and 9. Following this a double precision NumPy array is created using this data using the \texttt{PyArray\_SimpleNewFromData} command (lines 12-13). Note how \texttt{NPY\_DOUBLE} is specified as the data type and the array has dimensions \texttt{\{num\_cells,1\}} where the first dimension refers to the number of degrees of freedom present on this particular rank. Lines 15-20 use the \texttt{PyTuple\_SetItem} function to set the arguments for calling a function. In this case, the first argument is the NumPy array that was just created and the second dimension is the integer value of the current rank. The second argument is required for the purpose of book-keeping from within the Python module. Finally, the function is called with the specified tuple of arguments in line 23 using \texttt{PyObject\_CallObject}. The return value is cast to a pointer to a \texttt{PyArrayObject} and stored in \texttt{pValue}. Now we detail the process of receiving data from the Python functions (usually in the form of NumPy arrays) and interfacing it with the OpenFOAM data structure. This may be beneficial for in-situ computations of quantities that require Python packages, but are then utilized in the classical partial differential equation computation of OpenFOAM. Examples include the hybridized use of machine learning for bypassing one portion of the entire numerical solution. Line 25 to 37 detail how one may use \texttt{PyArray\_GETPTR2} to move data from the return value (a two-dimensional numpy array) to OpenFOAM's native data structure. This allows for preserving connectivity information when writing the results of this analysis to disk. Thus, classical visualization tools interfaced with OpenFOAM (such as Paraview) may be used for visualizing Python results.

\begin{lstlisting}[language=C++,style=mystyle,caption=Passing OpenFOAM data to NumPy through Python C-API and retrieving return value. Here we assume the x-component of velocity is sent to Python function which returns a pointwise computation, label=LST2]
/* Extract double precision data from OpenFOAM datastructure */
volScalarField ux_ = U.component(vector::X);

/* Storing data in double precision array for sending to Python */
forAll(ux_.internalField(), id) 
{   /* Storing velocities */
    input_vals[id][0] = ux_[id];
}

/* The dimensions of the numpy array to be created */
npy_intp dim[] = {num_cells, 1};
/* Pass data to numpy array */
array_2d = PyArray_SimpleNewFromData(2, dim, NPY_DOUBLE, &input_vals[0]);
    
/* Pass the array to the python function */
PyTuple_SetItem(python_func_args, 0, array_2d);

/* Pass the rank to the function */
rank_val = PyLong_FromLong(rank);
PyTuple_SetItem(python_func_args, 1, rank_val);

/* Call the function - pValue stores the return value of the function */
pValue = (PyArrayObject*)PyObject_CallObject(python_func, python_func_args); 

/* Return data from Python into OpenFOAM datastructure - allocate memory*/
volScalarField out_ = U.component(vector::X);

/* Load return value from Python into OpenFOAM data structure*/
for (int mode = 0; mode < truncation; ++mode)
{
    /* Overwriting data */
    forAll(out_.internalField(), id)
    {
        // Here we assume that pValue has a numpy array of dimension 2 with 1 column only
        out_[id] = *((double*)PyArray_GETPTR2(pValue, id, 0));
    }
}
\end{lstlisting}

%% file: Source/Compiling_Linking.tex
We briefly go over how the compiling and linking to Python and NumPy are performed using \texttt{wmake}, OpenFOAM's build system. A sample configuration file is shown in Listing \ref{Linking} which shows how one must add the paths to the various header files for the Python and Numpy APIs in the \texttt{EXE\_INC} field (lines 10 and 11). Similarly, paths to shared objects and link flags must also be provided in the \texttt{EXE\_LIB} field (line 13 and 25). It is particularly careful to avoid missing the right linking flags as a new solver may be compiled successfully but crash at runtime. 

\begin{lstlisting}[language=bash,style=mystyle,caption=OpenFOAM solver configuration for linking to Python.,label=Linking]
EXE_INC = \
    -I$(LIB_SRC)/MomentumTransportModels/momentumTransportModels/lnInclude \
    -I$(LIB_SRC)/MomentumTransportModels/incompressible/lnInclude \
    -I$(LIB_SRC)/transportModels/lnInclude \
    -I$(LIB_SRC)/finiteVolume/lnInclude \
    -I$(LIB_SRC)/sampling/lnInclude \
    -I$(LIB_SRC)/dynamicFvMesh/lnInclude \
    -I$(LIB_SRC)/dynamicMesh/lnInclude \
    -I$(LIB_SRC)/meshTools/lnInclude \
    -I/path_to_python_virtual_environment/include/python3.6m/ \
    -I/path_to_python_virtual_environment/lib/python3.6/site-packages/numpy/core/include \

EXE_LIBS = \
    -L/path_tolibpython/ \
    -lmomentumTransportModels \
    -lincompressibleMomentumTransportModels \
    -lincompressibleTransportModels \
    -lfiniteVolume \
    -lfvOptions \
    -lsampling \
    -ldynamicFvMesh \
    -ltopoChangerFvMesh \
    -ldynamicMesh \
    -lmeshTools \
    -lpython3.6m
\end{lstlisting}

%% file: Source/Algorithms.tex
\subsection{Online Singular Value Decomposition}

For demonstrating our tool, we will first use Levy and Lindenbaums method for performing a streaming SVD \cite{levy1998sequential}  in-situ. Our target application is to use this SVD for analyzing the presence of coherent structures in the flow field. Usually, this analysis is performed by constructing a data matrix $A \in \mathbb{R}^{M \times N}$. $N$ refers to the number of `snapshots' of data collected for the analysis and $M$ is the number of degrees of freedom in each snapshot. For the purpose of this analysis, $M >> N$ and the regular SVD gives us
\begin{align}
    A = U \Sigma V^T
\end{align}
where $U \in \mathbb{R}^{M \times N}, \Sigma \in \mathbb{R}^{N \times N}, V \in \mathbb{R}^{N \times N}$. The classical SVD computation scales as $O(MN^2)$ and requires $O(MN)$ memory. This analysis becomes intractable for computational physics applications such as CFD where the degrees of freedom may grow very large for simulating high wavenumber content. In their seminal paper, Levy and Lindenbaum proposed a streaming variant of the SVD that reduces the computational and memory complexity significantly. It performs this by extracting solely the first $K$ left singular vectors, which correspond to the $K$ largest coherent structures. Consequently, we are able to reduce the cost of the SVD to $O(MNK)$ operations and the memory footprint also reduces to $O(MK)$. This technique also has a streaming component to it, where the left singular eigenvectors may be updated in a batch-like manner. We summarize the procedure in Algorithm \ref{Algo1} of the Appendix.  A Python and NumPy implementation for this algorithm and how it may interface with OpenFOAM is shown in Listing \ref{LST4}. We remark that the scalar forget factor $ff$ (line 11), set between 0 and 1, controls the effect of older data batches on the final result for $U_i$. Setting this value to 1.0 implies that the online-SVD convergences to the regular SVD utilizing all the snapshots in one-shot. Setting values of $ff$ less than one reduces the impact of the snapshots observed in previous batches of the past. We utilize an $ff=0.95$ for this study. Specific examples that have been integrated with a novel OpenFOAM solver are available in our supporting repository \texttt{https://github.com/argonne-lcf/PythonFOAM}.

\begin{lstlisting}[language=python,style=mystyle,caption=Python module for online SVD update according to Levy and Lindenbaum \cite{levy1998sequential}.,label=LST4]
import numpy as np

class online_svd_calculator(object):
    """
    K : Number of modes to truncate
    ff : Forget factor
    """
    def __init__(self, K, ff):
        super(online_svd_calculator, self).__init__()
        self.K = K
        self.ff = ff

    def initialize(self,A):
        # Algorithm 1, Step I1
        q, r = np.linalg.qr(A)
        
        # Algorithm 1, Step I2
        ui, self.di, _ = np.linalg.svd(r)
        
        self.ui = np.matmul(q,ui)[:,:self.K]
        self.di = self.di[:self.K]

    def incorporate_data(self,A):
        """
        A is the new data matrix
        """
        # Algorithm 1, Step 1
        m_ap = self.ff*np.matmul(self.ui,np.diag(self.di))
        m_ap = np.concatenate((m_ap,A),axis=-1)
        udashi, ddashi = np.linalg.qr(m_ap)
        
        # Algorithm 1, Step 2
        utildei, dtildei, vtildeti = np.linalg.svd(ddashi)

        # Algorithm 1, Step 3
        self.di = dtildei[:self.K]
        utildei = utildei[:,:self.K]
        
        # Algorithm 1, Step 4
        self.ui = np.matmul(udashi,utildei)

def snapshot_func(array,rank):

    global iter, u_snapshots, v_snapshots, w_snapshots

    if iter == 0:
        print('Collecting snapshots iteration: ',iter)
        
        u_snapshots = array[:,0].reshape(-1,1)
        v_snapshots = array[:,1].reshape(-1,1)
        w_snapshots = array[:,2].reshape(-1,1)

        iter+=1
    else:
        print('Collecting snapshots iteration: ',iter)
        
        u_temp = array[:,0].reshape(-1,1)
        v_temp = array[:,1].reshape(-1,1)
        w_temp = array[:,2].reshape(-1,1)

        # Preallocation is possible to avoid array concatenation overhead
        u_snapshots = np.concatenate((u_snapshots,u_temp),axis=-1)
        v_snapshots = np.concatenate((v_snapshots,v_temp),axis=-1)
        w_snapshots = np.concatenate((w_snapshots,w_temp),axis=-1)

        iter+=1

    return 0

def svd_func(rank):
    
    global online_mode
    global iter, u_snapshots, v_snapshots, w_snapshots
    global init_mode, u_svd_calc, v_svd_calc, w_svd_calc

    if init_mode:
        print('Performing online SVD on snapshots rankwise - initialization')
        u_svd_calc.initialize(u_snapshots)
        v_svd_calc.initialize(v_snapshots)
        w_svd_calc.initialize(w_snapshots)

        init_mode = False
    else:
        u_svd_calc.incorporate_data(u_snapshots)
        v_svd_calc.incorporate_data(v_snapshots)
        w_svd_calc.incorporate_data(w_snapshots)

    u_modes = u_svd_calc.ui
    v_modes = v_svd_calc.ui
    w_modes = w_svd_calc.ui

    print('Modal calculation completed')
  
    u_snapshots = None
    v_snapshots = None
    w_snapshots = None
    iter = 0
    
    return_data = np.concatenate((u_modes,v_modes,w_modes),axis=0)

    return return_data
    
iter = 0
u_snapshots = None
v_snapshots = None
w_snapshots = None

online_mode = True
init_mode = True
u_svd_calc = online_svd_calculator(5,0.95)
v_svd_calc = online_svd_calculator(5,0.95)
w_svd_calc = online_svd_calculator(5,0.95)

\end{lstlisting}

\subsection{Distributed singular value decomposition}

In this section, we will introduce the approximate partitioned method of snapshots (APMOS) for computing distributed left singular vectors for our provided test cases. Note that the primary difference from the Online SVD is that this algorithm does not provide for a batch-wise update of the singular vectors. Instead, each batch has its respective basis vector calculation which is stored in an OpenFOAM compatible data structure to disk. While this algorithm loses the ability to construct a set of bases for the entire duration of the simulation, its distributed nature allows for the construction of a global basis even in the presence of a domain decomposition. This parallelized computation of the SVD was introduced in \cite{wang2016approximate} and we recall its main algorithm below. First, APMOS relies on the local calculation of the left singular vectors for the data matrix on each rank of the simulation. To construct this data matrix, snapshots of the local data may be collected over multiple timesteps. Each row of this matrix corresponds to a particular grid point and each column corresponds to a snapshot of data at one time instant. The first stage of local operations is thus
\begin{align}
    A^i = U^i \Sigma^i V^{*i}
\end{align}
where $i$ refers to the index of the rank ranging from 1 to $N_r$ (the total number of ranks), $U^i \in \mathbb{R}^{M_i \times N}$, $\Sigma^i \in \mathbb{R}^{N \times N}$, and $V^i \in \mathbb{R}^{N \times N}$. Here, $M_i$ refers to the number of grid points in rank $i$ of the distributed simulation. Note that instead of an SVD, one may also perform a method of snapshots approach for computing $V^i$ at each rank provided $M_i >> N$. A column-truncated subset of the right singular vectors, $\tilde{V}^i$, and the singular values, $\tilde{\Sigma}^i$, may then be sent to one rank to perform the exchange of global information for computing the POD basis vectors. This is obtained by collecting the following matrix at rank 0 using the MPI gather command
\begin{align}
    W = \left[ \tilde{V}^1 (\tilde{\Sigma}^1)^T, ..., \tilde{V}^{N_r} (\tilde{\Sigma}^{N_r})^T \right].
\end{align}
In this study, we utilize a truncation factor $r_1=$50 columns of $V_i$ and $\Sigma_i$ for broadcasting. Subsequently a singular value decomposition of $W$ is performed to obtain
\begin{align}
    W = X \Lambda Y^*. 
\end{align}
Given another threshold factor $r_2$ corresponding to the number of columns retained for $X$, a reduced matrix $\tilde{X}$ and reduced singular values $\tilde{\Lambda}$ is broadcast to all ranks. The distributed \emph{global} left singular vectors may then be assembled at each rank as follows for each basis vector $j$
\begin{align}
    \tilde{U}_j^i = \frac{1}{\tilde{\Lambda}_j} A^i \tilde{X}_j
\end{align}
where $\tilde{U}_j^i$ is the $j^{\text{th}}$ singular vector in the $i^{\text{th}}$ rank, $\tilde{\Lambda}_j$ is the $j^{\text{th}}$ singular value and $\tilde{X}_j$ is the $j^{\text{th}}$ column of the reduced matrix $\tilde{X}$. In this study, we choose $r_2=5$ columns for our threshold factor for this last stage. We note that the choices for $r_1$ and $r_2$ may be used to balance communication costs and accuracy for this algorithm. Pseudocode \ref{Algo2}, in the appendix, summarizes this procedure. Listing \ref{LST5} details the Python implementation for this algorithm to interface with OpenFOAM.

\begin{lstlisting}[language=python,style=mystyle,caption=Python module for approximate partitioned method of snapshots for left singular vector computation \cite{wang2016approximate}.,label=LST5]
import numpy as np
import mpi4py
mpi4py.rc.initialize = False
mpi4py.rc.finalize = False
from mpi4py import MPI

iter = 0
u_snapshots = None
v_snapshots = None
w_snapshots = None
num_modes = 5 # This should match truncation

def snapshot_func(array,rank):

    global iter, u_snapshots, v_snapshots, w_snapshots

    if iter == 0:
        print('Collecting snapshots iteration: ',iter)
        
        u_snapshots = array[:,0].reshape(-1,1)
        v_snapshots = array[:,1].reshape(-1,1)
        w_snapshots = array[:,2].reshape(-1,1)

        iter+=1
    else:
        print('Collecting snapshots iteration: ',iter)
        
        u_temp = array[:,0].reshape(-1,1)
        v_temp = array[:,1].reshape(-1,1)
        w_temp = array[:,2].reshape(-1,1)

        u_snapshots = np.concatenate((u_snapshots,u_temp),axis=-1)
        v_snapshots = np.concatenate((v_snapshots,v_temp),axis=-1)
        w_snapshots = np.concatenate((w_snapshots,w_temp),axis=-1)

        iter+=1

    return 0

# Method of snapshots to accelerate
def generate_right_vectors_mos(Y):
    new_mat = np.matmul(np.transpose(Y),Y)
    w, v = np.linalg.eig(new_mat)

    svals = np.sqrt(np.abs(w[:rval]))
    
    # Threshold r1 
    rval = 50

    return v[:,:rval].astype('double'), svals.astype('double') 

def apmos_func(placeholder):
    
    global iter, u_snapshots, v_snapshots, w_snapshots # Iteration and local data

    comm = MPI.COMM_WORLD
    rank = comm.Get_rank()
    nprocs = comm.Get_size()

    snapshots_list = [u_snapshots, v_snapshots, w_snapshots]
    phi_list = []

    for local_data in snapshots_list:
        local_data_mean = np.mean(local_data,axis=1)
        local_data = local_data-local_data_mean[:,None]

        # Run a method of snapshots
        vlocal, slocal = generate_right_vectors_mos(local_data)

        # Find W
        wlocal = np.matmul(vlocal,np.diag(slocal).T)

        # Gather data at rank 0:
        wglobal = comm.gather(wlocal,root=0)

        # perform SVD at rank 0:
        if rank == 0:
            temp = wglobal[0]
            for i in range(nprocs-1):
                temp = np.concatenate((temp,wglobal[i+1]),axis=-1)
            wglobal = temp

            x, s, y = np.linalg.svd(wglobal)
            
            # Truncation threshold r2
            rval = num_modes
            
            x = x[:,:rval]
            s = s[:rval]
        else:
            x = None
            s = None
        
        x = comm.bcast(x,root=0)
        s = comm.bcast(s,root=0)

        # perform singular vector calculation at each local rank
        phi_local = []
        for mode in range(rval):
            phi_temp = 1.0/s[mode]*np.matmul(local_data,x[:,mode:mode+1])
            phi_local.append(phi_temp)

        temp = phi_local[0]
        for i in range(rval-1):
            temp = np.concatenate((temp,phi_local[i+1]),axis=-1)
        
        phi_list.append(temp)

    # Clean memory
    u_snapshots = None
    v_snapshots = None
    w_snapshots = None

    iter = 0
    return_data = np.concatenate((phi_list[0],phi_list[1],phi_list[2]),axis=0)
    return return_data
\end{lstlisting}

\subsection{Nonlinear compression using deep autoencoder}

\begin{figure}
    \centering
    \includegraphics[width=0.9\textwidth]{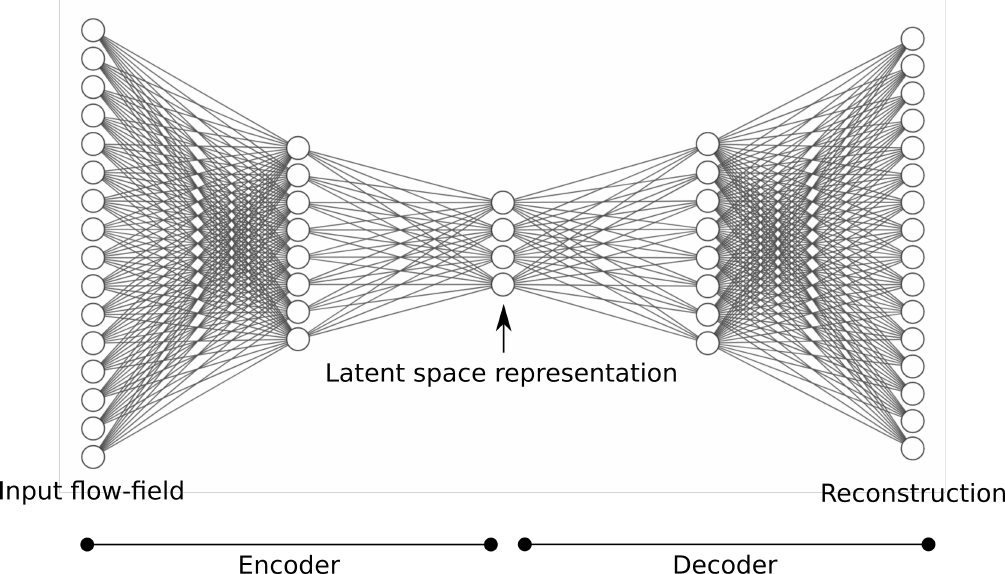}
    \caption{A representative schematic of the deep neural network autoencoder. A bottleneck architecture is used while learning an identity function to obtain a low-dimensional embedding of the flow-field.}
    \label{fig:auto_schem}
\end{figure}

Now we demonstrate an application where a deep learning architecture is utilized to find a nonlinear low-dimensional embedding from several snapshots of transient data. Autoencoders have been successfully used for various reduced-order modeling, data-compression, and data exploration problems \cite{maulik2021reduced,hasegawa2020cnn,murata2020nonlinear,nakamura2021convolutional,gonzalez2018deep}. Classical workflows for training autoencoders are similar to performing singular value decompositions, and involve storing several snapshots to disk at predefined temporal checkpoints. Subsequently, a separate computational workflow is executed to train the deep learning architecture, usually on specialty hardware. In this study, we deploy a deep neural network autoencoder to use a snapshot from each iteration of the PimpleFOAM solver and obtain a latent space representation. A representative schematic of the fully-connected autoencoder architecture is shown in Figure \ref{fig:auto_schem}. We use the Swish activation function at each layer, a batch size of 128 snapshots, the ADAM optimizer with a learning rate of 0.001, and more importantly, a bottleneck width of 4 neurons for this architecture. This means that the trained encoder of the autoencoder can compress the flow-field to a four-dimensional state and reconstruct from the same. A TensorFlow model definition of the autoencoder is also given in Listing \ref{LST6}. We clarify that the implementation of the deep neural network autoencoder is performed on the same resource as that used for the numerical solve and at periodic intervals when adequate training data has been collected. Also, while concurrent training of autoencoder and data collection is possible, we leave that to a future implementation. 

\begin{lstlisting}[language=python,style=mystyle,caption=A deep neural network autoencoder used for compressing flow-field information,label=LST6]
# Encoder
encoder_inputs = Input(shape=(num_points*num_fields),name='Field')

x = Dense(50, activation=my_swish)(encoder_inputs)
x = Dense(25, activation=my_swish)(x)
x = Dense(10, activation=my_swish)(x)
encoded = Dense(self.num_latent)(x)
self.encoder = Model(inputs=encoder_inputs,outputs=encoded)
    
# Decoder
decoder_inputs = Input(shape=(self.num_latent,),name='decoded')
x = Dense(10,activation=my_swish)(decoder_inputs)
x = Dense(25,activation=my_swish)(x)
x = Dense(50,activation=my_swish)(x)
decoded = Dense(num_points*num_fields)(x)

self.decoder = Model(inputs=decoder_inputs,outputs=decoded)

# Autoencoder
ae_outputs = self.decoder(self.encoder(encoder_inputs))
self.model = Model(inputs=encoder_inputs,outputs=ae_outputs,name='Autoencoder')

weights_filepath = 'weights.h5'
my_adam = optimizers.Adam(lr=0.001, beta_1=0.9, beta_2=0.999, epsilon=None, decay=0.0, amsgrad=False)
checkpoint = ModelCheckpoint(weights_filepath, monitor='val_loss', verbose=1, save_best_only=True, mode='min',save_weights_only=True)
earlystopping = EarlyStopping(monitor='val_loss', min_delta=0, patience=100, verbose=0, mode='auto', baseline=None, restore_best_weights=True)
callbacks_list = [checkpoint,earlystopping]

# Compile network
self.model.compile(optimizer=my_adam,loss='mean_squared_error')    
self.model.summary()
\end{lstlisting}

%% file: Source/Experiments.tex
\textcolor{black}{In this section, we shall outline several experiments that demonstrate the utility of our Python bindings to OpenFOAM. We shall discuss results of an in-situ data analysis using the previously introduced online-SVD, distributed SVD, and the deep learning autoencoder. Scaling analyses on different architectures will also be provided for the distributed SVD. }

All our experiments are performed with the PimpleFOAM solver in OpenFOAM. PimpleFOAM is an unsteady incompresible solver which allows for turbulence-scale resolving Large Eddy simulations (LES), Reynolds average Navier Stokes (RANS) simulations or hybrid RANS/LES simulations. The solver also uses a merged PISO (Pressure Implicit with Splitting of Operator) and SIMPLE (Semi-Implicit Method for Pressure-Linked Equations) algorithm (PIMPLE) for the velocity-pressure coupling for which an inner PISO iteration is performed to get an initial solution which is then corrected using an outer SIMPLE iteration.  The PIMPLE algorithm provides for enhanced stability of the solver. In particular, we are able to utilize larger time-steps and, hence, larger Courant numbers that are greater than unity. This is an advantage over the PISO algorithm which is restricted by a stable Courant number criterion of less then unity \cite{holzmann2016mathematics}. 
At the end of each PimpleFOAM solver update, our Python bindings are called to either send snapshots to a NumPy array for collection or to perform a data analysis operation (such as an SVD computation or a neural network training). For timesteps during which the analyses are performed, postprocessed data are also returned to OpenFOAM which uses its native I/O operations to write to disk. This enables the use of Paraview for visualization as is usual.\textcolor{black}{ We note that our implementation is independent of the numerical discretization or solver algorithm. An overall workflow for our deployments is as follows}:

\begin{algorithm}[h!]
\SetAlgoLined
\KwResult{Postprocessed data (Singular vectors, autoencoder reconstructions)}
 Initialize solver, Python/C API coupling\;
 \While{While not final time}{
  Perform solver integration\;
  Cast OpenFOAM data to NumPy array\;
  \eIf{Data analysis checkpoint}{
   Retrieve data from NumPy array\;
   Perform data analyses (deep learning/SVD)\;
   Send postprocessed data back to OpenFOAM\;
   Write data to disk using OpenFOAM I/O\;
   }{
   Collect data in NumPy arrays\;
  }
 }
 \caption{A prototypical PythonFOAM coupling workflow for the various experiments in this article.}
\end{algorithm}

\subsection{LES of backward facing step}
\label{bfs_les}

\begin{figure}
    \centering
    \includegraphics[width=\textwidth]{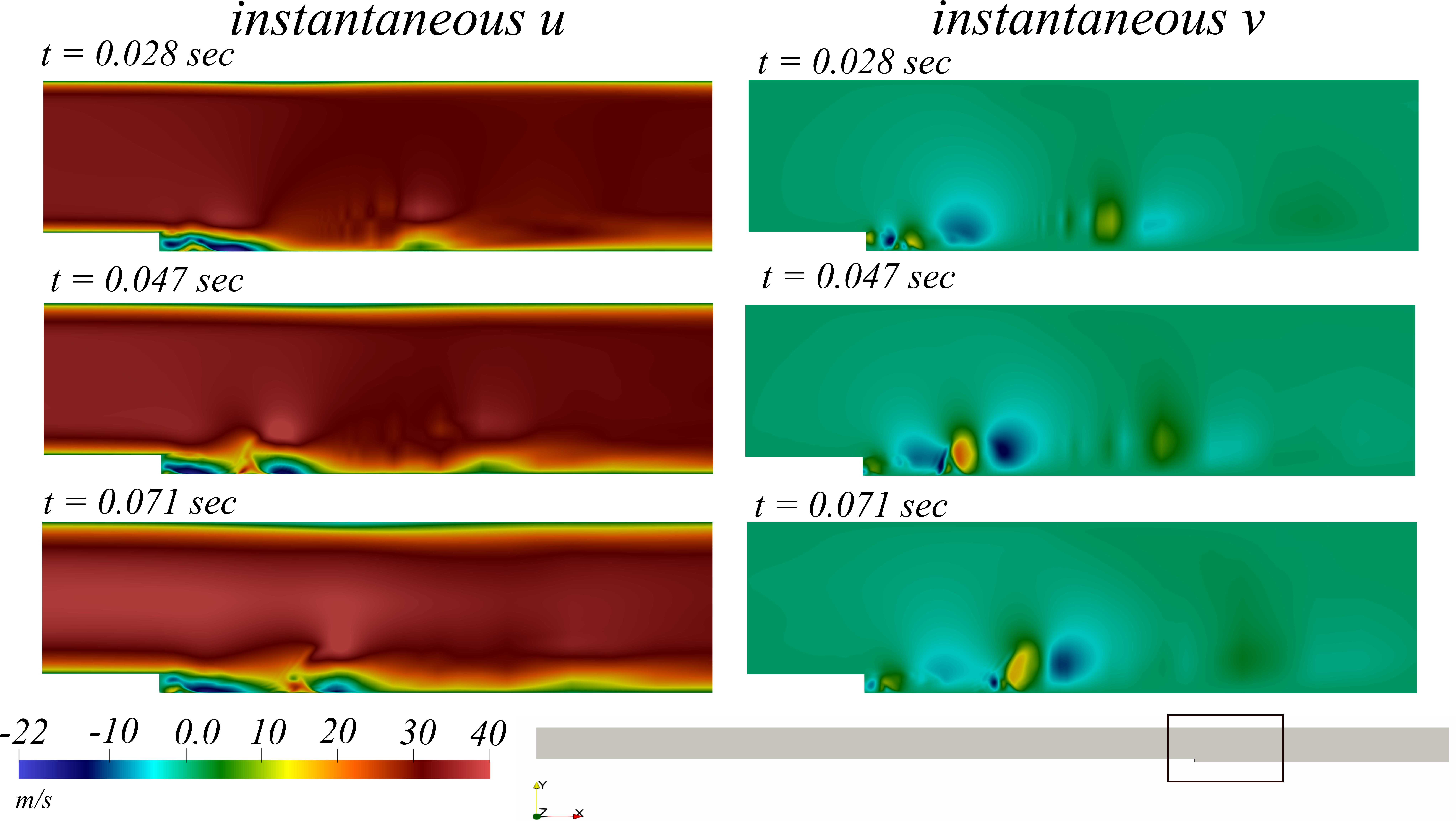}
    \caption{Dimensionless instantaneous streamwise and vertical velocities $u$ and $v$ at three characteristic time instances.}
    \label{fig:velocities}
\end{figure}

\begin{figure}
    \centering
    \includegraphics[width=\textwidth]{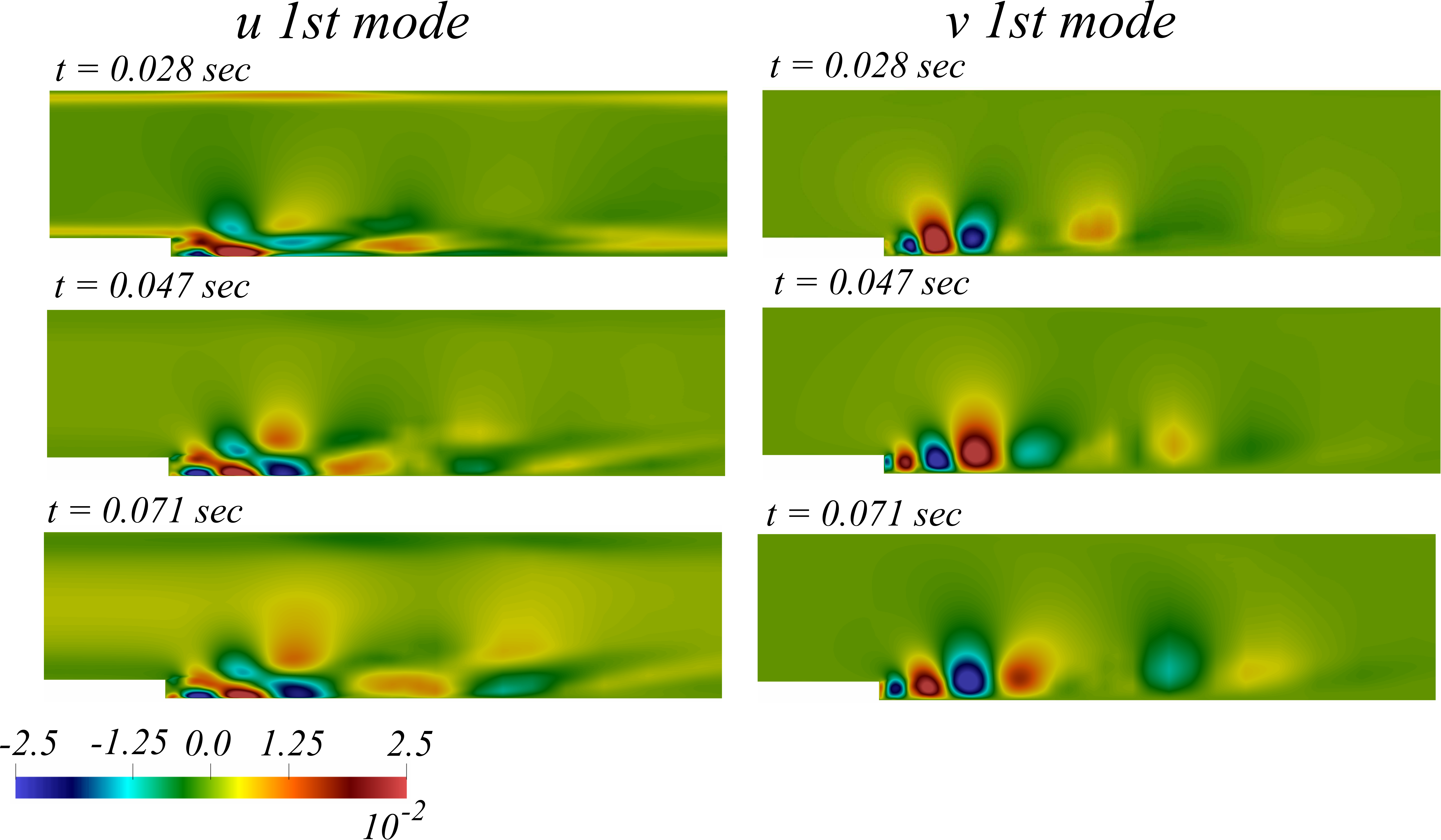}
    \caption{First singular vectors for $u$ and $v$ at three characteristic time instances.}
    \label{fig:1stMode}
\end{figure}

\begin{figure}
    \centering
    \includegraphics[width=\textwidth]{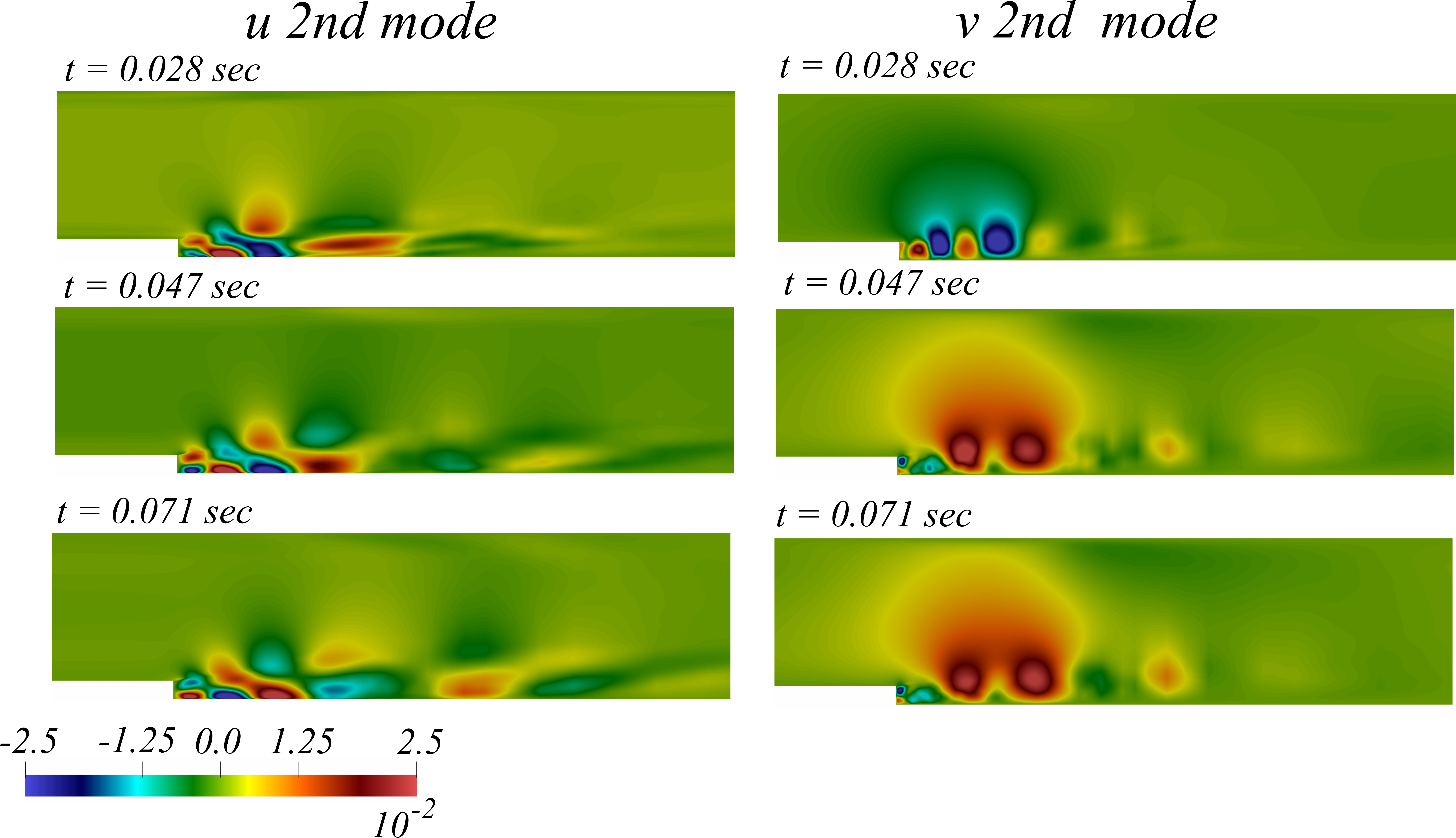}
    \caption{Second singular vectors for $u$ and $v$ at three characteristic time instances.}
    \label{fig:2ndMode}
\end{figure}

Our first experiment is that of the large-eddy simulation (LES) of a two-dimensional backward facing step using the standard Smagorinsky implementation in OpenFOAM. We note that the two-dimensional assessment does not truly qualify as a physical case (since LES requires a three-dimensional domain). However, for this assessment, we merely wish to perform an in-situ data analysis using the online-SVD and compare with the instantaneous flow features obtained from the numerical methods. The purpose of this computational assessment to confirm that the online-SVD is able to detect coherent structures in the flow field. The left singular vectors of this SVD correspond to structures in the flow field where there are concentrations of kinetic energy. This can be seen by comparing the instantaneous velocity contours of the $x$ and $y$ components of the velocity and the structures observed when overlapping the singular vectors on to the computational mesh. The singular vectors for both $u$ and $v$ show structures emanating downstream of the step where shedding and recirculation are present. These singular vectors may also be used to represent the flow field in a low-dimensional form through forming a subspace from a limited number of orthonormal basis vectors. However, that study is not explored in this work. Figure \ref{fig:velocities} shows instantaneous snapshots for the $x$ and $y$ components of velocity for the backward facing step exhibiting unsteady separation and shedding behavior downstream of the step. This behavior is successfully captured in the singular vectors shown in Figures \ref{fig:1stMode} and \ref{fig:2ndMode}. Coherent structures in the singular vectors of the $y$ component of the velocity depict the oscillatory nature of the shedding as well.

\begin{figure}
    \centering
    \includegraphics[width=\textwidth]{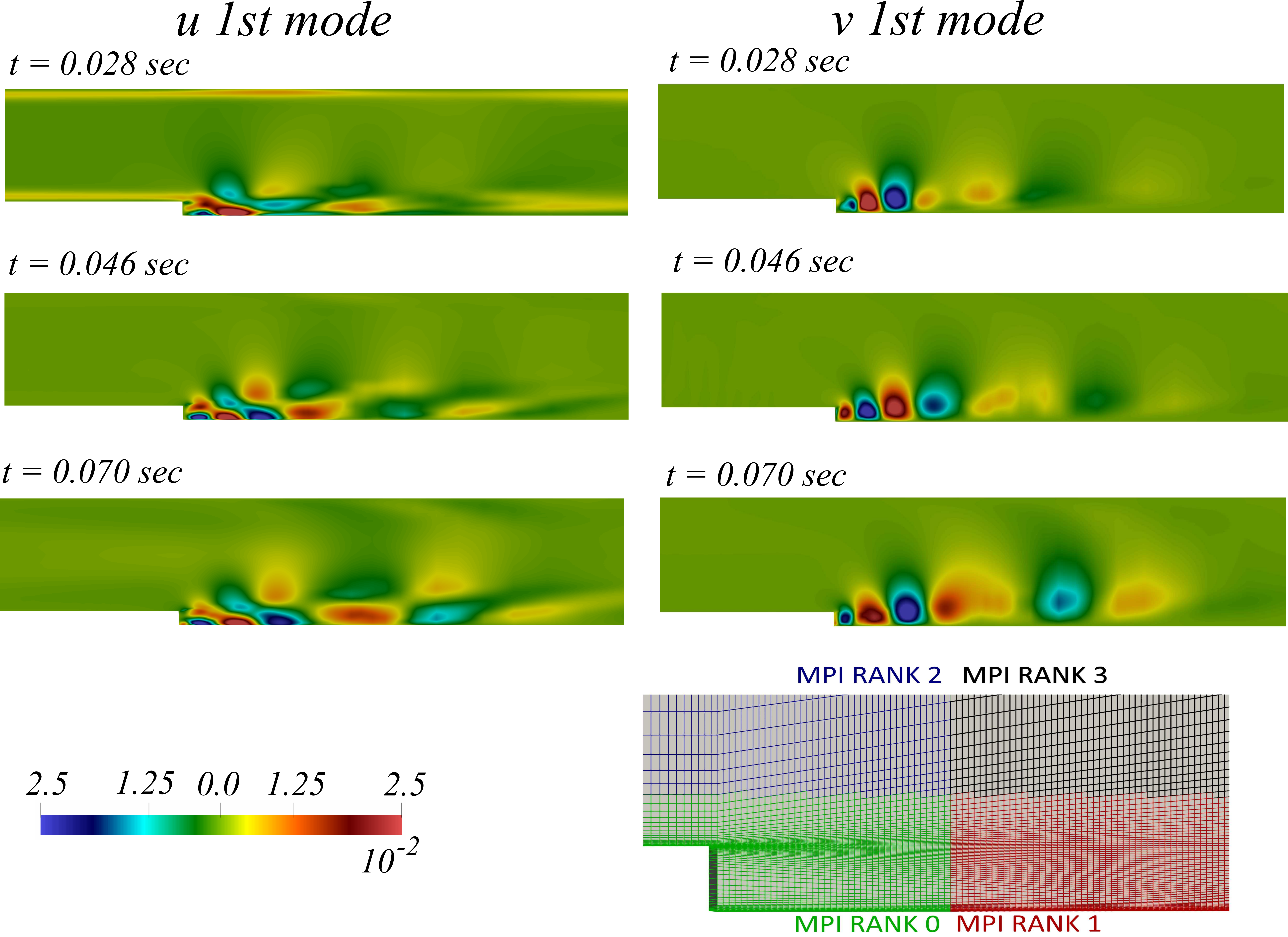}
    \caption{First singular vectors for $u$ and $v$ at three characteristic time instances using the APMOS method. Note how information from 4 ranks is used to construct the global basis vectors.}
    \label{fig:1stMode_APMOS}
\end{figure}

\begin{figure}
    \centering
    \includegraphics[width=\textwidth]{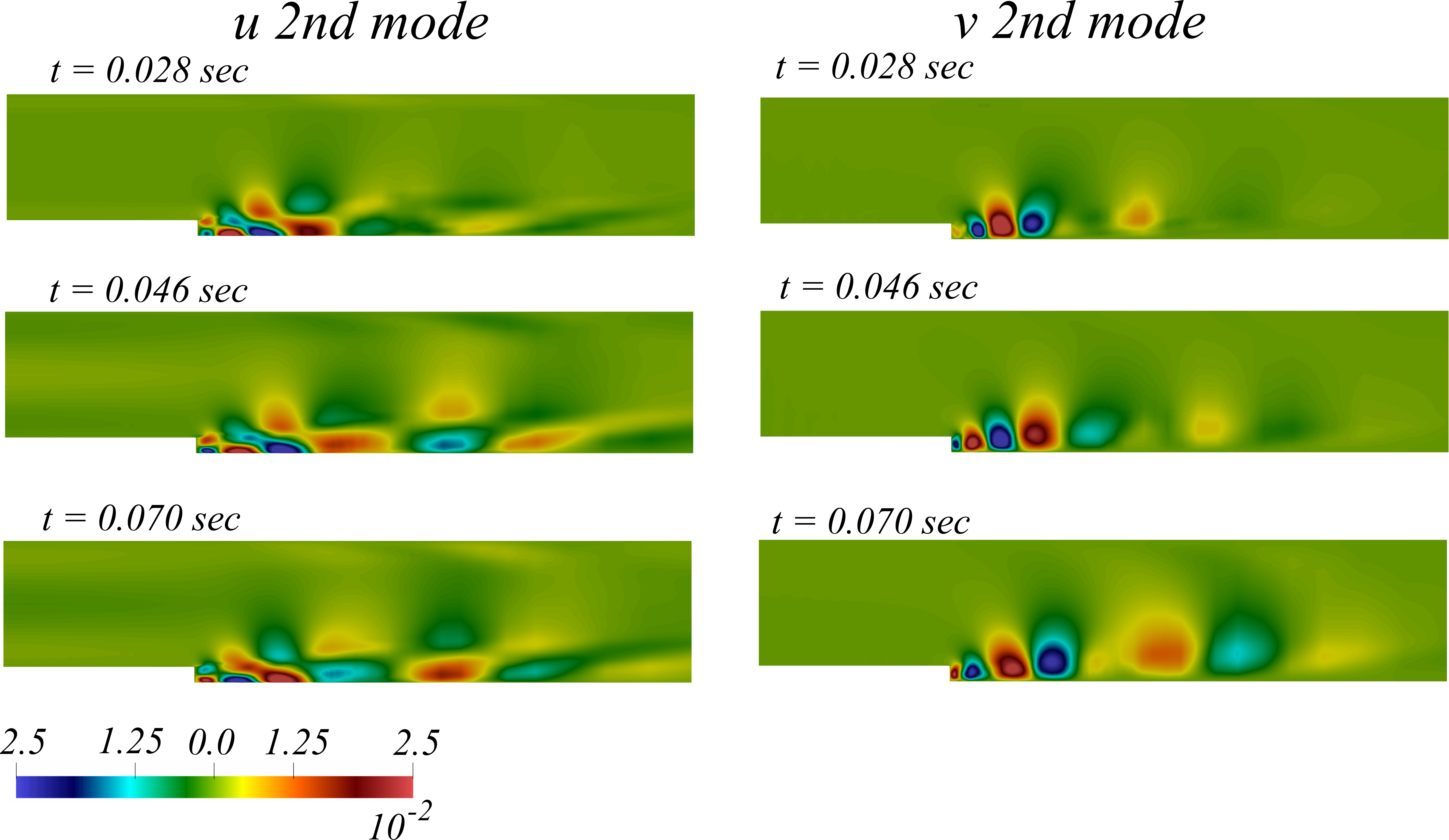}
    \caption{Second singular vectors for $u$ and $v$ at three characteristic time instances using the APMOS method. Note how information from 4 ranks is used to construct the global basis vectors.}
    \label{fig:2ndMode_APMOS}
\end{figure}

Figure \ref{fig:1stMode_APMOS} shows a visualization indicating how the backward facing step was distributed across 4 ranks and how APMOS is able to approximate the global left singular vectors effectively. Figure \ref{fig:2ndMode_APMOS} outlines the second singular vectors obtained using APMOS. In this experiment, snapshots were collected and utilized for the distributed SVD every 2000 iterations of PimpleFOAM. One can observe that the singular vectors correspond to the coherent structures in the flow-field in a continuous sense, despite the distributed nature of the computation. We remark here that differences may be observed with the online-SVD results from the previous set of experiments which may be attributed to factors such as the additional level of truncation employed in the global communication of local right singular vectors (see step I2 and I3 in Algorithm \ref{Algo2}). However, the basis vectors obtained through APMOS adequately reveal coherent structures in the flow and are orthonormal, which allows for their use in downstream tasks such as projection-based modeling or analyses. An important capability is highlighted here - the utilization of one MPI communicator across two languages C++ and Python. While the MPI communication is initiated in OpenFOAM, the Python MPI4PY library is able to send and receive NumPy array data between Python interpreters residing at different ranks. This enables possibilities for the use of distributed algorithms such as data-parallel machine learning in future extensions. 

Subsequently, we analyse the scaling of our distributed SVD in the following. We evaluate computational efficiency for two different processing architectures which are commonly available in modern Petascale systems. Our computational set up is given by a well-known OpenFOAM tutorial case: the large eddy simulation of a turbulent channel flow at friction Reynolds number 395 using the WALE turbulence model. We perform strong scaling assessments on a grid with 3.84 million degrees of freedom on 1, 4, 8, 16, 32, 64, and 128 compute nodes of Bebop at the Laboratory Computing Resource Center (LCRC) of Argonne National Laboratory. Cells were split across ranks using the Scotch decomposition technique \cite{pellegrini1996scotch} to ensure appropriate load-balancing. Our cell-counts for distributed experiments were  approximately 960,000 cells per rank for a 4 rank simulation, 480,000 cells per rank of a 8 rank simulation, 240,000 cells per rank for a 16 rank simulation, 120,000 cells per rank for a 32 rank simulation,  60,000 cells per rank for a 64 rank simulation, and 30,000 cells per rank for a 128 rank simulation.

As mentioned previously, these experiments are for two different computer architectures: Intel's Knights Landing (KNL) and Broadwell (BDW). Each BDW node (Xeon E5-2695v4) comes with 36 cores, 128 GB of DDR4 memory and maximum memory bandwidth of 76.8 GB/s \cite{IntelBDW}. BDW also supports the Advanced Vector Extensions (AVX2) instruction set which is capable of 256-bit wide SIMD operations. In contrast, each KNL node (Xeon Phi 7230) has almost twice the core-count with 64 cores and 96 GB of DDR4 memory. The KNL comes with the AVX-512 instruction set which has 512-bit wide SIMD operations. Each KNL also comes with MCDRAM which is a high-bandwidth memory that is integrated on-package with a total of 16 GB space; for the STREAM TRIAD benchmark, the bandwidth coming from the MCDRAM has been reported to be over 450 GB/s \cite{jeffers2016intel}. 


The total computational cost at each time step may be decomposed into several sub-costs as outlined in the following. We have costs associated with the partial differential equation update (i.e., the PimpleFOAM solver update), the time taken to record snapshots, the time taken to perform the distributed SVD, the time taken to copy singular vector data into OpenFOAM data structures, the I/O time to write out the flow field information, and the time required to cast data from the OpenFOAM data structures to NumPy arrays. \textcolor{black}{Note that OpenFOAM function objects allow for runtime and post-hoc data analyses without typecasts but we choose to construct our wrapper through the pathway of a novel solver construction for the ease of downstream modification for arbitrary tasks. In principle, function objects that utilize Python/C coupling via NumPy arrays may be constructed and used in concurrence with the demonstrated procedure.} Averaged times for the aforementioned operations are recorded over the duration of a simulation across multiple ranks and shown in addition to strong scaling plots. We note that SVD computation, data copy, and IO computation times are obtained by averaging across only five such instances for each simulation. In contrast, other computations such as the cost for PDE solution update, snapshot collection, and NumPy casting were averaged across all the time steps for each simulation. 

\begin{figure}
    \centering
    \mbox{
    \subfigure[Cost per timestep]{\includegraphics[width=0.49
    \textwidth]{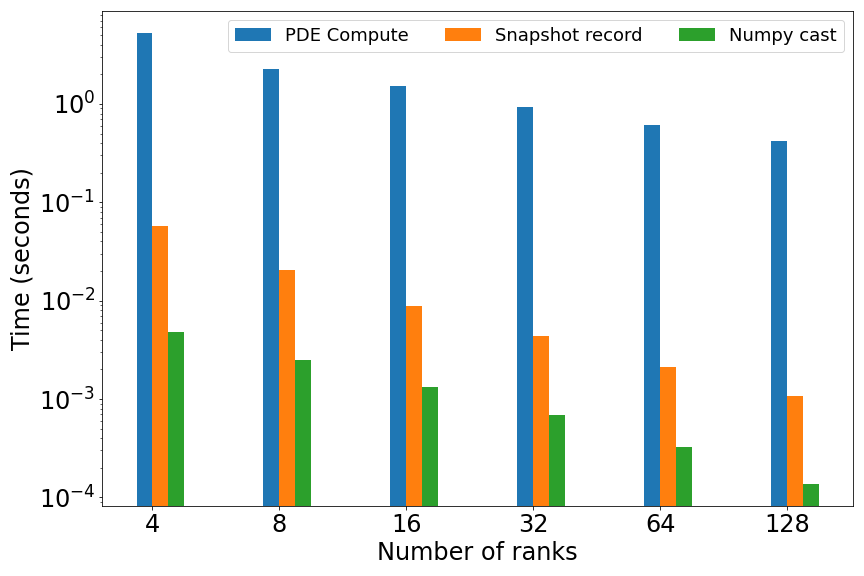}}
    \subfigure[Cost per 2000 timesteps]{\includegraphics[width=0.49
    \textwidth]{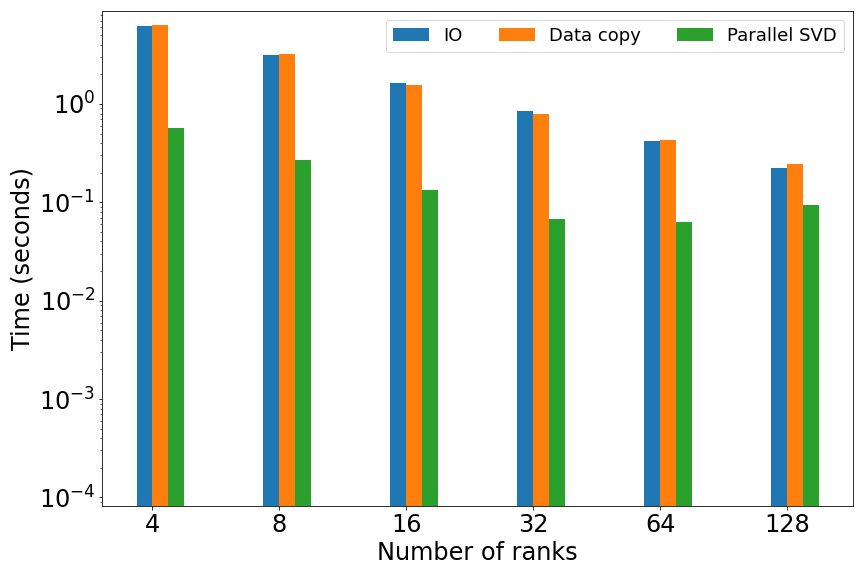}}
    } \\
    \mbox{
    \subfigure[Scaling]{\includegraphics[width=0.5
    \textwidth]{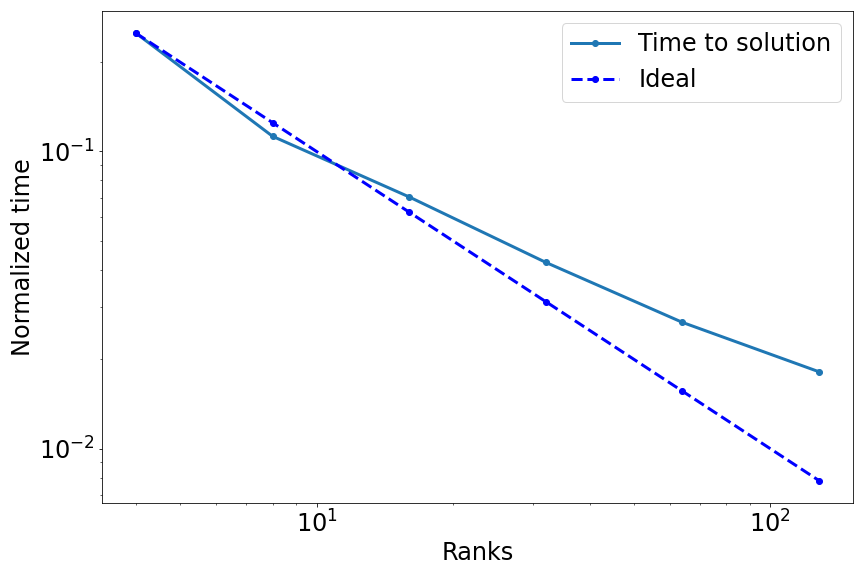}}
    }
    \caption{Scaling analyses of PythonFOAM for an in-situ distributed singular value decomposition on multiple ranks of the LCRC machine Bebop (with the Intel Broadwell architecture). Figure (a) shows the breakdown of costs incurred at each time step. Figure (b) shows the breakdown of costs incurred every 2000 timesteps. Figure (c) shows strong scaling for the total time to solution.}
    \label{fig:bdw_scaling_apmos}
\end{figure}

\begin{figure}
    \centering
    \mbox{
    \subfigure[Cost per timestep]{\includegraphics[width=0.49
    \textwidth]{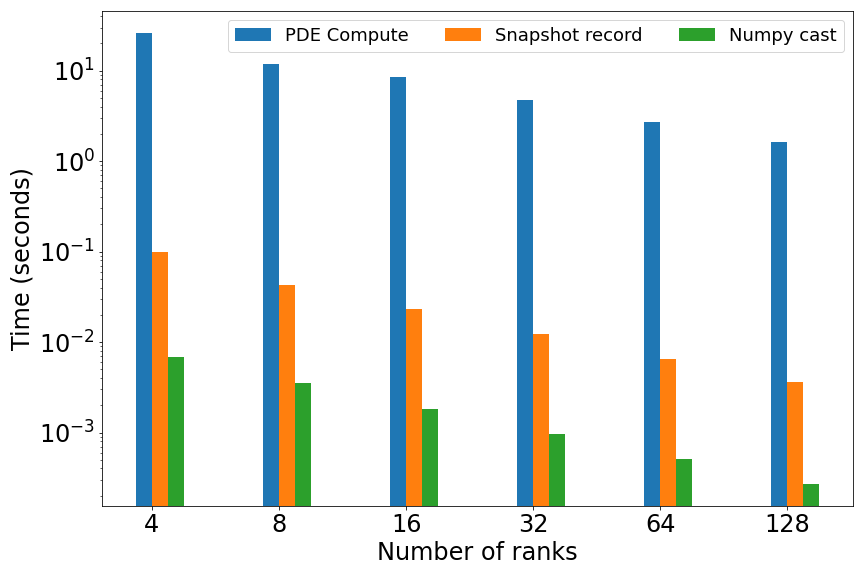}}
    \subfigure[Cost per 2000 timesteps]{\includegraphics[width=0.49
    \textwidth]{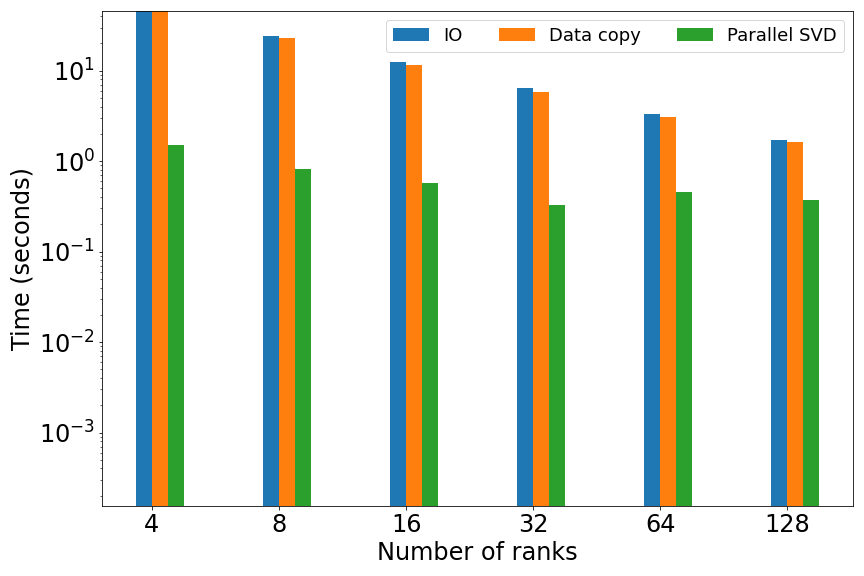}}
    } \\
    \mbox{
    \subfigure[Scaling]{\includegraphics[width=0.5
    \textwidth]{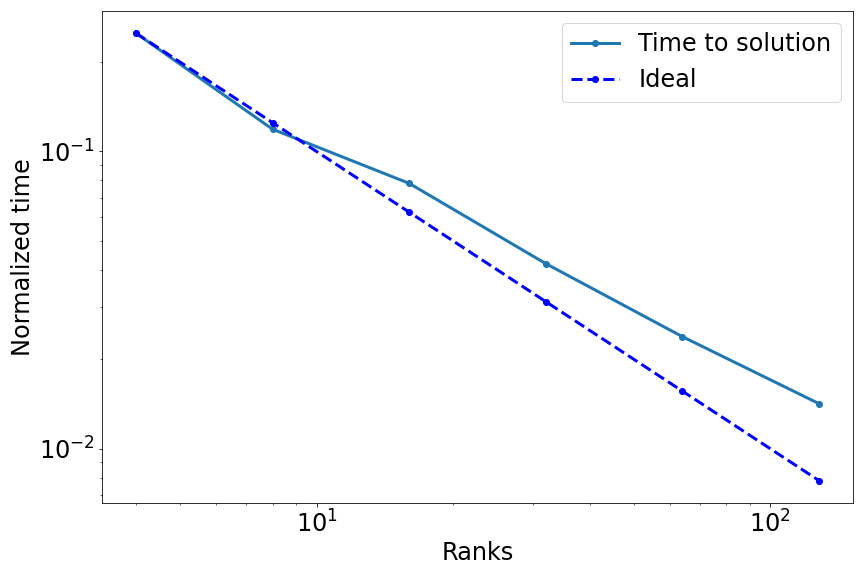}}
    }
    \caption{Scaling analyses of PythonFOAM for an in-situ distributed singular value decomposition on multiple ranks of the LCRC machine Bebop (with the Intel Knights-Landing architecture).  Figure (a) shows the breakdown of costs incurred at each time step. Figure (b) shows the breakdown of costs incurred every 2000 timesteps. Figure (c) shows strong scaling for the total time to solution.}
    \label{fig:knl_scaling_apmos}
\end{figure}

Strong scaling results for the distributed SVD are presented in Figures \ref{fig:bdw_scaling_apmos} and \ref{fig:knl_scaling_apmos} for the Broadwell and KNL architectures respectively. The x-axis in these plots shows the ranks while the y-axis indicates the time required to perform a given operation (e.g. PimpleFOAM compute time, SVD time, etc.). The cost breakdowns provided with each plot show how computation associated with the numerical solver significantly dominates other costs (since this cost is incurred at each time step of the numerical simulation). It must be noted here that for utilizations of the OpenFOAM-Python coupling where data must be sent \emph{and} received at frequent intervals, the data copy time may prove to be a potential bottleneck that adds costs on the order of the PDE-compute itself. An example is the use of a machine learning algorithm that predicts a transient flow-field quantity. This must be accounted for prior to designing data-driven solutions to classical problems in scientific computing such as for deep learning surrogates for turbulence modeling. We also remark that, for the implementation tested here, performance on Broadwell nodes is remarkably faster than Knights-Landing. However, we note that several optimization strategies may be deployed for the latter that can improve performance considerably.

\subsection{Deep learning autoencoder}

\begin{figure}
    \centering
    \mbox{
    \includegraphics[width=0.48\textwidth]{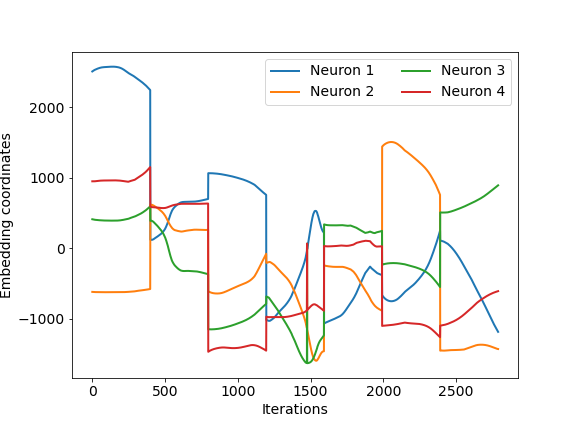}
    \includegraphics[width=0.48\textwidth]{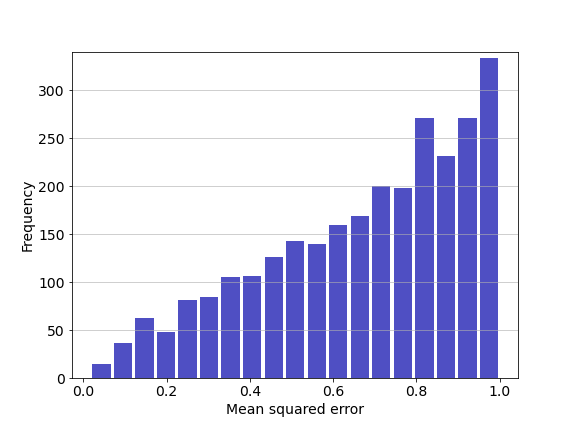}
    }
    \caption{Evolution of backward facing step in encoded coordinates as obtained by training several deep neural network autoencoders at intervals of 400 iterations (left) and histogram of reconstruction snapshot mean squared errors (right). Note that a new set of coordinates are obtained at the end of each 400 iteration period of full-order snapshot collection. }
    \label{fig:encoded_evolution}
\end{figure}

\begin{figure}
    \centering
    \includegraphics[width=\textwidth]{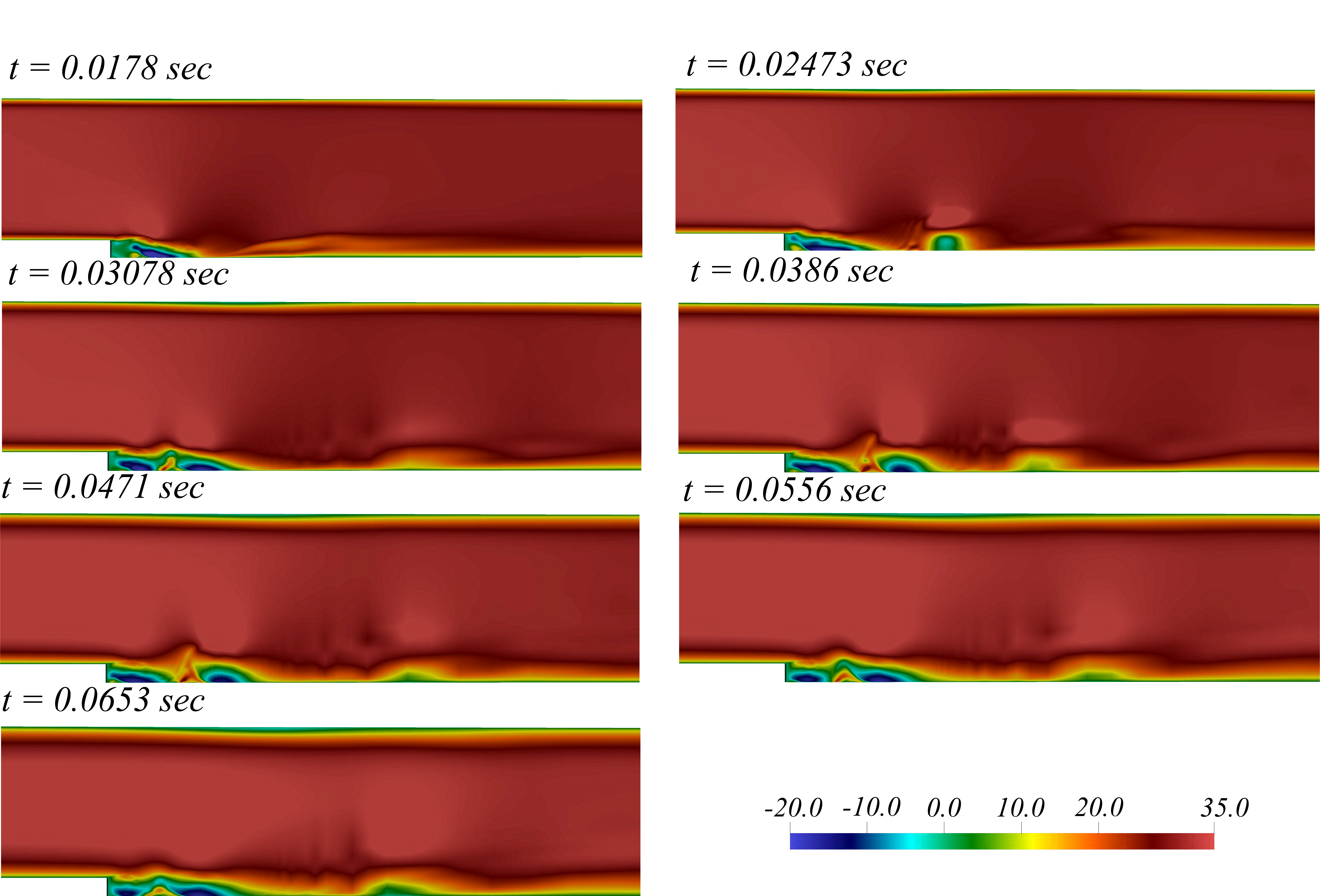}
    \caption{Streamwise velocity component $u$, obtained using numerical solver PimpleFOAM.}
    \label{fig:u_cfd}
\end{figure}

\begin{figure}
    \centering
    \includegraphics[width=\textwidth]{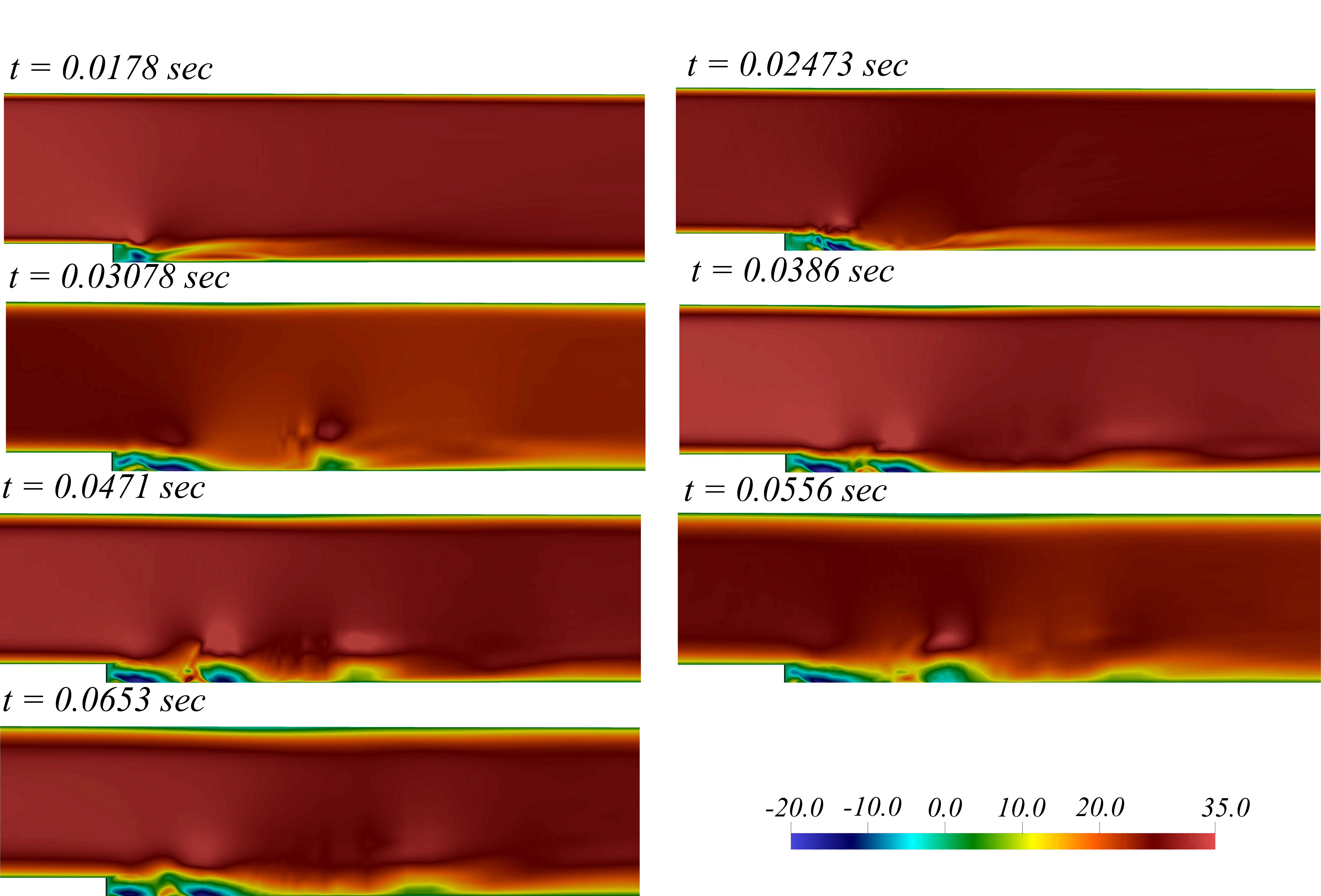}
    \caption{Streamwise velocity component $u$, obtained using deep neural network autoencoder reconstructions from low-dimensional embedding in TensorFlow.}
    \label{fig:u_rec}
\end{figure}

\textcolor{black}{Our final set of experiments demonstrates the use of a deep neural network autoencoder for compressing flow-field information without having to checkpoint data to disk. For this experiment, we use snapshots of the $x-$component of velocity that are collected into sets of 400 numerical iterations and subsequently used to train the deep neural network. The trained network weights are saved to disk at the end of 400 iterations and the learned model is subsequently used to obtain low-dimensional representations of the flow-field snapshots for the next set of 400 iterations (which are also collected for training the next autoencoder). We remark here, for clarity, that a unique autoencoder is trained at the end of each collection of 400 snapshots. In practice, algorithmic augmentations could be added such as carrying over the previous best model and retraining through transfer learning. The evolution of our backward facing step (introduced in Section \ref{bfs_les}), in the encoded coordinates, is shown in Figure \ref{fig:encoded_evolution}. We remark here that the discontinuities at the end of each 400-iteration window correspond to a novel set of coordinates having been obtained by a new network training. These low-dimensional representations may then be used to reconstruct, approximately, the high dimensional flow field using the decoder network of the appropriately trained autoencoder. Figure \ref{fig:encoded_evolution} also shows the reconstruction root-mean-squared errors for the various test snapshots. While the original magnitude of the flow-field is comparable to the inlet velocity (i.e., 32.2 m/s), these errors are an order of magnitude lower. Examples of the CFD flow fields obtained by the numerical solver for this experiment are shown in Figure \ref{fig:u_cfd} and their corresponding reconstructions from  the low-dimensional embedding can be seen  in Figure \ref{fig:u_rec}. We note that these reconstructions are approximate - and retrieved from a nonlinear transformation of the low-dimensional embedding using the decoder of the autoencoder deep neural network.}

%% file: Source/Conclusions.tex
This article outlines results, and provides source codes, for an extension of OpenFOAM that allows for an arbitrary interface with Python. Our interface is built on the Python/C API and allows for the utilization of Python modules, class objects, and functions in addition to the exchange of data between OpenFOAM and Python via the NumPy/C API. This tool can therefore precipitate a wide variety of in-situ data analytics, visualization, and machine learning for computational fluid dynamics applications within a reproducible open source environment.

Our interface is demonstrated through the deployment of a streaming singular value decomposition on a transient canonical backward facing step problem that exhibits shedding and separation downstream of the step. The singular vectors obtained from the streaming approach can be used to detect the presence of coherent structures for similar problems. Our example shows how one may extract these singular vectors by calling a snapshot record function and a singular value decomposition function during the deployment of a transient OpenFOAM solver (pimpleFOAM). The singular vectors may then be stored within OpenFOAM itself to preserve connectivity information and to utilize integrated visualization with Paraview. Furthermore, we also deploy the proposed wrapper on a canonical channel flow problem at $Re_\tau=395$ to assess the parallel efficiency of the OpenFOAM/Python interface. We do this on different computing architectures for a distributed SVD based on the approximate partitioned method of snapshots. Scaling analyses of these experiments indicate scaling efficiency in the presence of distributed data analyses may be preserved if the computational cost of the solver dominates that of the data storage and typecast costs at each time step. This has implications for the use of distributed data science frameworks in concurrence with PDE solvers at extreme scales. \textcolor{black}{Finally, we also demonstrate a state-of-the-art use case where a deep neural network with a bottleneck architecture is used to learn low-dimensional representations of the flow-field. This network, an autoencoder, is used to track the evolution of dynamics on the embedding coordinates and provide approximate reconstructions from this space on demand through decoding. This proves that arbitrary deep learning tasks may be deployed within OpenFOAM using the Python API of TensorFlow.}

Our extensions to this library shall investigate asynchronous couplings with specialized computational resources to mitigate the aforementioned in-situ data analytics bottleneck at extreme scales. The series of experiments in this article firmly fall within the umbrella of tightly coupled analyses, where PDE computation and analyses are performed on the same computational resource in a blocking procedure. In other words, until data analysis is complete, PDE compute is halted and vice versa. Our future goals are to loosen this coupling on heterogeneous HPC systems, where data analysis may be performed on devoted accelerators while PDE computation is performed on simulation-friendly devices. An asynchronous interface (i.e., a `loose-coupling') between the two would allow for data analysis and simulation to be performed concurrently and for improvements in scaling characteristics. Our current and future investigations are structured along the aforementioned verticals.

%% file: Source/Appendix.tex
\begin{algorithm}[H]
\label{Algo1}
\SetAlgoLined
\KwResult{The truncated left singular vector matrix $U_i$ after $i$ batch iterations.}
 \textbf{Parameters:} \\
 Forget factor $ff$\;
 \textbf{Initialization:} \\
 Initial data matrix $A_0 \in \mathbb{R}^{M \times B}$ where $B$ is the number of snapshots per batch\;
 I1. Perform QR-decomposition : $A_0=QR$ \;
 I2. Perform SVD of $R=U'D_0 V_0^T$ and obtain $U_0 = Q U'$ \;
 \While{New data $A_i$ available}{
  1. Compute QR decomposition after column-wise concatenation of new data: $\left[ff \cdot U_{i-1} D_{i-1} \mid A_{i}\right]=U_{i-1}^{\prime} D_{i-1}^{\prime}$\;
  2. Compute SVD of $D_{i-1}' = \tilde{U}_{i-1} \tilde{D}_{i-1} \tilde{V}^{T}$\;
  3. Preserve the first $K$ columns of $\tilde{U}_{i-1}$ and denote $\hat{U}_{i-1}$ \;
  4. Obtain the updated left singular vectors: $U_i = U_{i-1}' \hat{U}_{i-1}$ \;
  5. Truncate to retain K values of $\tilde{D}_{i-1}$ to obtain $D_i$ \;
 }
 \caption{Streaming singular value decomposition \cite{levy1998sequential}.}
\end{algorithm}

\begin{algorithm}[H]
\label{Algo2}
\SetAlgoLined
\KwResult{The truncated left singular vector matrix $\tilde{U}^i$ in each rank $i$ of a distributed computation}
 \textbf{Parameters:} \\
 Threshold factors $r_1$ and $r_2$\;
 \textbf{Algorithm:} \\
 Local data matrix $A^i$ at each rank $i$ of distributed simulation\;
 1. Perform local SVD or method of snapshots calculation of local right singular vectors : $A^i = U^i \Sigma^i V^{*i} $. \;
 2. Truncate $V^i$, $\Sigma^i$ by retaining only $r_1$ columns to obtain $\tilde{V}^i$ and $\tilde{\Sigma}^i$ respectively. \;
 3. Obtain $W = \left[ \tilde{V}^1 (\tilde{\Sigma}^1)^T, ..., \tilde{V}^{N_r} (\tilde{\Sigma}^{N_r})^T \right]$ at rank 0 using MPI Gather.\;
 4. Perform SVD, $W=X \Lambda Y^*$ at rank 0.\;
 5. Truncate $X, \Lambda$ by retaining only $r_2$ columns to obtain $\tilde{X}, \tilde{\Lambda}$ respectively.\;
 6. Send $\tilde{X}, \tilde{\Lambda}$ to each rank using MPI Broadcast.\;
 7. Obtain local partition of $j^{\text{th}}$ global left singular vector $\tilde{U}_j^i = \frac{1}{\tilde{\Lambda}_j} A^i \tilde{X}_j$ where $j$ corresponds to the column of the respective matrices.
\caption{Distributed singular value decomposition \cite{wang2016approximate}}.
\end{algorithm}